\documentclass[A4]{article}

\usepackage[utf8]{inputenc}
\usepackage[english]{babel}
\usepackage{graphicx}
\usepackage{subfigure}
\usepackage{comment}
\usepackage{amsmath}
\usepackage{booktabs}
\usepackage{authblk}
\usepackage{color}

\def\beq{\begin{equation}}
\def\endeq{\end{equation}}

\usepackage[usenames,dvipsnames,svgnames,table]{xcolor}
\def\Frac#1#2{\frac{\displaystyle #1}{\displaystyle #2}}

\numberwithin{equation}{section}

\renewenvironment{remark}{}{}
\def\Frac#1#2{\frac{\displaystyle #1}{\displaystyle #2}}

\date{}

\title{The adjoint neutron transport equation and the statistical approach for its solution}

\author[1]{P.~Saracco}
\author[2]{S.~Dulla}
\author[2]{P.~Ravetto} 
\affil[1]{\small{INFN- Sezione di Genova \\ Via Dodecaneso, 33 - 16146 Genova (Italy)}}
\affil[2]{\small{Politecnico di Torino, Dipartimento Energia \\ Corso Duca degli Abruzzi, 24 - 10129 Torino (Italy)}}

\begin{document}

\maketitle

\begin{abstract}
The adjoint equation was introduced in the early days of neutron transport and its solution, the neutron importance, has ben used for several applications in neutronics. The work presents at first a critical review of the adjoint neutron transport equation. Afterwards, the adjont model is constructed for a reference physical situation, for which an analytical approach is viable, i.e. an infinite homogeneous scattering medium. This problem leads to an equation that is the adjoint of the slowing-down equation that is well-known in nuclear reactor physics. A general closed-form analytical solution to such adjoint equation is obtained by a procedure that can be used also to derive the classical Placzek functions. This solution constitutes a benchmark for any statistical or numerical approach to the adjoint equation. 
A sampling technique to evaluate the adjoint flux for the transport equation is then proposed and physically interpreted as a transport model for pseudo-particles. This can be done by introducing appropriate kernels describing the transfer of the pseudo-particles in phase space. This technique allows estimating the importance function by a standard Monte Carlo approach.
The sampling scheme is validated by comparison with the analytical results previously obtained. 
\end{abstract}


\section{Introduction}

The introduction of the adjoint neutron transport was one of the key landmarks in the evolution of nuclear reactor physics. The solution of this equation was interpreted as neutron importance and opened the way to many applications in nuclear reactor physics and engineering. The basic concept of neutron importance was established by Weinberg and Wigner \cite{wigner}, although the idea was proposed in various forms also by other Authors. A consistent derivation of the adjoint Boltzmann equation for the critical reactor in steady state and its physical interpretation as the neutron importance conservation equation in integro-differential form is due to Ussachoff \cite{ussa}. The integral form of the adjoint transport equation has also been introduced \cite{robkin} and in a recent contribution the connection between adjoints and Green's functions is highlighted \cite{imre}.

The concept was also generalized to source-driven systems and to time-dependent situations. A general approach to the theory of neutron importance was proposed and discussed by Lewins \cite{lewins}, as "the physical basis of variational and perturbation theory in transport and diffusion problems". As this subtitle of Lewins's book clearly states, the theory of the adjoint function lays the foundation for the applications of perturbation and variational methods in the field of nuclear reactor physics. Over the years, these methods have provided powerful and effective tools for the analysis of nuclear reactors. The literature on the interpretation and on the applications of the adjoint function is huge (it is impossible to give an exhaustive list of references here; see the bibliography in Lewins's book that covers at least the earliest history \cite{lewins}).

A significant thrust forward in perturbation analysis is due to Gandini \cite{gandini1,gandini2}. His generalizations led to a huge extension of the possibilities of the perturbative approach in various fields of applied sciences. As an example, in nuclear reactor physics, the technique could be effectively used in the fields of nuclide evolution and fuel cycle \cite{gandini3} and it can be applied also to non-linear problems. A theory of the adjoint function can be developed also for source-driven problems, once a problem-tailored definition of the adjoint function is introduced \cite{cadinu}.

The methods developed for nuclear reactor kinetics rely heavily on the neutron importance concept. The standard kinetic equations for the point reactor were consistently derived by projection of the neutron balance equations on the adjoint function \cite{henry}. The various quasi-static schemes for spatial and spectral kinetics nowadays used for time-dependent full-core simulations are based on this idea (see, for instance, \cite{mund}). At last it is worth to cite the use of the adjoint quantities in modern sensitivity analysis and uncertainty quantification, which have a crucial role in today's nuclear science \cite{hany}.

Some efforts were made in the past to sample weighted quantities directly in the Monte Carlo (MC) process. The significant work by Rief on direct perturbation evaluations by Monte Carlo should be acknowledged \cite{rief1, rief2}. Also some more recent work must be acknowledged \cite{Serpent1}. The sampling of weighted quantities has been introduced for the evaluation of integral reactor parameters \cite{kiedro,Serpent,Tripoli}.

The capability to evaluate the adjoint function is included in the standard deterministic neutronic codes used for reactor analysis. Monte Carlo statistical methods are gaining a prominent role in a wide set of nuclear applications and the information on neutron importance is being used to guide the sampling procedure and speed up its convergence. The importance sampling techniques can be included within the frame of the so-called contributon theory \cite{dubi}. Several works have been performed with the objective of accelerating the statistical convergence of Monte Carlo (see, for instance, \cite{larsen} and the bibliography therein). However, the possibility of using Monte Carlo for the solution of the adjoint equation is very attractive per se. 
Two approaches are possible: either a backward neutron propagation technique \cite{hoog} or a proper forward procedure \cite{Tripoli}.
Many Authors have tackled the problem, with various attempts to maintain the same sampling approach as the one used in the direct Monte Carlo simulation, although no physical interpretation of such procedures is usually given \cite{carter, hoog, feghi, diop}. The procedure leads to the introduction of a concept of pseudo-particles, named adjunctons, which, through an appropriate transport process, are distributed as the adjoint function. In this framework, starting from the integral form of the transport equation, the work by Irving is certainly standing \cite{irving}.

In the present work a consistent approach to the sampling procedure for a Monte Carlo simulation for the adjoint function is illustrated and a physical interpretation is discussed. The procedure draws its inspiration from the work carried out by De Matteis \cite{dematteis}.  Further developments were presented in a later work  \cite{dematteis1}. The concept of pseudo-particles named adjunctons and of adjoint cross sections was used in these works, as well as in the works by Eriksson \cite{Erik} and by Irving \cite{irving} 

In the following the sampling procedure for the solution of the adjoint equation is defined for a fixed source problem, although it could be easily extended to eigenvalue simulations. The interest is mainly focused on the energy and direction variables, since the problem in space can be handled by simple extension of the procedure for the direct equation. Afterwards, the validation of the sampling procedure is considered. A reliable benchmark can be employed for this purpose: analytical benchmarks are particularly useful for a sound validation, since they are not affected by any discretization or truncation error and they have been widely proposed for various physical problems in transport theory \cite{ganapol}. To obtain an analytical benchmark the classical problem of neutron slowing down in an infinite medium is considered. The direct problem leads to the classical Placzek functions \cite{plac}. On the other hand, in this work, the adjoint equation is solved analytically using the same approach as for the direct equation, showing an interesting and useful duality property, and the results are compared to the ones obtained by Monte Carlo. As a last outcome of the work here presented, a fully analytical closed-form for both the direct and adjoint Placzek functions is obtained.

\section{Sampling procedure for the adjoint of the neutron transport equation}

The neutron transport equation reads \cite{duder}:
\begin{eqnarray}
\left(\Frac{1}{v}\Frac{\partial}{\partial t}+\hat\Omega\cdot\vec\nabla+\Sigma_t(\vec r,E)\right)\phi(\vec r, E,\hat\Omega,t)&=&
{\mathcal S}(\vec r, E,\hat\Omega,t)\\\nonumber && \hskip-4.5truecm
\int_{4\pi}d\hat\Omega'\int_0^\infty\,dE'\phi(\vec r, E',\hat\Omega',t)\Sigma(\vec r,E'\longrightarrow E,\hat\Omega'\longrightarrow\hat\Omega),
\label{eqn:xxxxyyy}
\end{eqnarray}
where, when necessary, also the fission process is introduced in the source term as:
\begin{eqnarray}
{\mathcal S}(\vec r, E,\hat\Omega,t)&=&{\mathcal S}_{\rm ext}(\vec r, E,\hat\Omega,t)+\\\nonumber &&
\hskip-2.5truecm +\Frac{\chi(E)}{4\pi}
\int_{4\pi}d\hat\Omega'\int_0^\infty\,dE'\phi(\vec r, E',\hat\Omega',t)\nu(\vec r,E')\Sigma_f(\vec r, E').
\end{eqnarray}

The static version of the transport equation is now considered. In the presence of fission and of an external source a physically meaningful solution exists, i.e. non-negative over the whole phase space considered, only if the fundamental multiplication eigenvalue $k$, defined by:
\begin{eqnarray}
\left(\hat\Omega\cdot\vec\nabla+\Sigma_t(\vec r,E)\right)\phi_k(\vec r,E,\hat\Omega) &=& \int_{4\pi}d\hat\Omega'\int_0^\infty\,dE'\phi_k(\vec r,E',\hat\Omega')\Sigma(\vec r,E'\longrightarrow E,\hat\Omega'\longrightarrow\hat\Omega)
\nonumber \\ && 
\hskip-2.5truecm +\frac{1}{k}\Frac{\chi(E)}{4\pi}
\int_{4\pi}d\hat\Omega'\int_0^\infty\,dE'\phi_k(\vec r,E',\hat\Omega')\nu(E')\Sigma_f(\vec r,E'),
\end{eqnarray}
is strictly smaller than unity ($k<1$).

For the purpose of the present work the (possible) space dependence is not relevant and, therefore, the static version of Eq. (\ref{eqn:xxxxyyy}) for an infinite homogeneous system with homogeneous isotropic source is considered:
\begin{eqnarray}
\Sigma_t(E)\phi(E,\hat\Omega) &=& {\mathcal S}_{\rm ext}(E)
+\int_{4\pi}d\hat\Omega'\int_0^\infty\,dE'\phi(E',\hat\Omega')\Sigma(E'\longrightarrow E,\hat\Omega'\longrightarrow\hat\Omega)
\nonumber \\ && +\Frac{\chi(E)}{4\pi}
\int_{4\pi}d\hat\Omega'\int_0^\infty\,dE'\phi(E',\hat\Omega')\nu(E')\Sigma_f(E').
\end{eqnarray}
Clearly, the angular flux must be space independent. Furthermore, if an isotropic medium is considered, the transfer kernel depends only on $\hat\Omega'\cdot\hat\Omega\equiv\cos\theta$ and, hence, the angular flux is independent of $\hat\Omega$. This is physically easily understandable, since the flux must be isotropic in an isotropic homogeneous infinite medium, since no source of anisotropy is present. It can be also proved mathematically, observing that, in such a case, the collision integral in the r.h.s. of the above equation, by integration over all $\hat\Omega'$, obviously turns out to be $\hat\Omega$ independent. The equation takes the following form:
\begin{eqnarray}
\label{eqn:directeq}
\Sigma_t(E)\phi(E) &=& {\mathcal S}_{\rm ext}(E)
+\int_{4\pi}d\hat\Omega'\int_0^\infty\,dE'\phi(E')\Sigma(E'\longrightarrow E,\hat\Omega'\cdot\hat\Omega)
\nonumber \\ && +\Frac{\chi(E)}{4\pi}
\int_{4\pi}d\hat\Omega'\int_0^\infty\,dE'\phi(E')\nu(E')\Sigma_f(E').
\end{eqnarray}
One can now define partial collision kernels through the following expression:
\begin{equation}
\begin{array}{l}
\displaystyle\int_{4\pi}d\hat\Omega'\int_0^\infty\,dE'\phi(E')\Sigma(E'\longrightarrow E,\hat\Omega'\cdot\hat\Omega)
\\+\displaystyle\Frac{\chi(E)}{4\pi}\int_{4\pi}d\hat\Omega'\int_0^\infty\,dE'\phi(E')\nu(E')\Sigma_f(E')
\\ \equiv \displaystyle\int_{4\pi}d\hat\Omega'\int_0^\infty\,dE'\phi(E')\Sigma_t(E')\displaystyle\sum_{k=s,f,\dots}\Frac{\Sigma_k(E')}{\Sigma_t(E')}\nu_k(E')f_k(E'\longrightarrow E,\hat\Omega'\cdot\hat\Omega),
\end{array}
\label{eqn:forwone}
\end{equation}
where $\nu_k(E)$ represents - in analogy with the usual fission term - the {\em mean number of neutrons emitted by a collision of type $k$
which has been triggered by a neutron of energy $E$}\footnote{In the r.h.s. of Eq. (\ref{eqn:forwone}) it is possible to include also collision processes other than fission or scattering.}. 
The {\em total collision kernel} $f(E'\longrightarrow E,\hat\Omega'\cdot\hat\Omega)$
is then  expressed as a probability-weighted sum of partial collision kernels:
\begin{equation}
f(E'\longrightarrow E,\hat\Omega'\cdot\hat\Omega)=
\sum_{k=s,f,\dots}\Frac{\Sigma_k(E')}{\Sigma_t(E')}\nu_k(E')f_k(E'\longrightarrow E,\hat\Omega'\cdot\hat\Omega).
\label{eqn:forwtwo}
\end{equation}
In these last two relations it is implicit that the energy-angle distributions  $f_k(E'\longrightarrow E,\hat\Omega'\cdot\hat\Omega)$ are normalized {\em with respect to the outgoing neutron energy $E$}, so that a probabilistic intepretation, useful for sampling in a Monte Carlo procedure, is natural: whenever a neutron with (incoming) energy $E'$ suffers a collision, which is of the kind $j$ with probability $\Frac{\Sigma_j(E')}{\Sigma_t(E')}$, then a mean number $\nu_j(E')$ of neutrons exits from collision with energy and angular distribution given by $f_j(E'\longrightarrow E,\hat\Omega'\cdot\hat\Omega)$. These relations are the conceptual basis for the Monte Carlo sampling process in neutron transport. As anticipated, the space dependence is omitted in the present discussion\footnote{It enters the simulation only through the determination of the next collision site, whose distance is ruled by total macroscopic cross section. This holds true also for the adjoint case, provided signs of velocities are reversed \cite{irving}.}.

By straighforward mathematical reasoning, the equation adjoint to (\ref{eqn:directeq})  takes the following form:
\begin{eqnarray}
\label{eqn:adj}
\Sigma_t(E)\phi^\dagger(E) &=& {\mathcal S}^\dagger_{\rm ext}(E)+\\ \nonumber && \hskip-2.5truecm 
+\int_{4\pi}d\hat\Omega'\int_0^\infty\,dE'\phi^\dagger(E')\Sigma_t(E)f(E\longrightarrow E',\hat\Omega'\cdot\hat\Omega).
\end{eqnarray}
A few comments on the physical meaning of this equation are worth-while. Although one refers to $\phi^\dagger$ as the "adjoint flux", physically it is not a flux. It is known as "neutron importance", it is not a density and as such it is a dimensionless quantity, quite differently from the neutron flux. This fact leads also to an interpretation of the integral terms in the above equation (\ref{eqn:adj}) that is quite different from the interpretation of the corresponding terms in equation (\ref{eqn:directeq}). For instance, to physically derive the scattering integral term in the balance established by Eq. (\ref{eqn:directeq}), one takes the total track length within the elementary volume $d\hat\Omega'dE'$, i.e. $\phi(E') d\hat\Omega'dE'$, and multiplies by the transfer function $\Sigma(E'\longrightarrow E,\hat\Omega'\cdot\hat\Omega)$, in order to obtain the number of neutrons emitted per unit energy and per unit solid angle at $E$ and $\hat\Omega$. The integration collects the contributions from all possible incoming energies and directions. On the other hand, for the balance of importance in Eq. (\ref{eqn:adj}), one must collect the contributions to importance of all neutrons generated by the scattering of a neutron characterized by energy $E$ and direction $\hat\Omega$. Therefore $\Sigma_t(E)f(E\longrightarrow E',\hat\Omega'\cdot\hat\Omega) d\hat\Omega' dE'$ is the fraction of scattered neutrons within $d\hat\Omega'dE'$ and, consequently, their contributions to the balance of importance is obtained multiplying by the importance of neutrons at the outgoing energy $E'$ and direction $\hat\Omega'$. The integration now collects the contributions from all possible outgoing energies and directions. 

The simplest way to obtain a basis for the MC simulation of the adjoint flux is to manipulate Eq. (\ref{eqn:adj}) in such a way as to obtain a set of relations formally identical to (\ref{eqn:forwone}, \ref{eqn:forwtwo}); we remark that the main difficulty in developing a sampling scheme for (\ref{eqn:adj}) stems from the fact that in this case $E'$ is the energy of particles outgoing from the collision. It is clear that this difficulty can be (formally) overcome by defining
\begin{equation}
\Sigma_t(E)f(E\longrightarrow E',\hat\Omega'\cdot\hat\Omega)=
\Sigma_t^\dagger(E')f^\dagger(E'\longrightarrow E,\hat\Omega'\cdot\hat\Omega)
\label{eqn:adjf}
\end{equation}
in such a way that (\ref{eqn:adjf}) appears identical to (\ref{eqn:forwone}), {\em provided one assumes - or better defines} -
$\Sigma_t^\dagger(E)=\Sigma_t(E)$, which implies that the total rate of collision for the pseudo-particles here implicitly introduced into the game\footnote{In literature we have two naming choices, pseudo-neutrons or adjunctons.} is the same as for the corresponding physical particles: this is the only physical constraint we assume to set up a simulation framework for the adjoint equation. In this way we obtain for the adjoint equation:
\begin{eqnarray}
\label{eqn:adj2}
\Sigma_t(E)\phi^\dagger(E) &=& {\mathcal S}^\dagger_{\rm ext}(E)+\\ \nonumber && \hskip-2.5truecm 
+\int_{4\pi}d\hat\Omega'\int_0^\infty\,dE'\phi^\dagger(E')\Sigma^\dagger_t(E')f^\dagger(E'\longrightarrow E,\hat\Omega'\cdot\hat\Omega)\,.
\end{eqnarray}
We underline that the superscript $\dagger$ does not imply here transposition and complex conjugation, but it simply hints to the fact
that dagged quantities refer to the parameters defining the transport properties of pseudo-particles: through this identification a purely formal transposition acquires a true physical meaning.
However this is not sufficient, because we must also require that the adjoint kernel takes the form of a sum of partial collision kernels for pseudo-particles, namely:
\begin{equation}
f^\dagger(E'\longrightarrow E,\hat\Omega'\cdot\hat\Omega)=
\sum_{k=s,f,\ldots}\Frac{\Sigma_k^\dagger(E')}{\Sigma_t^\dagger(E')}\nu_k^\dagger(E')
f_k^\dagger(E'\longrightarrow E,\hat\Omega'\cdot\hat\Omega),
\label{111}
\end{equation}
so that we can interpret all the dagged quantities in the same fashion as the original macroscopic cross sections for neutrons, in particular the fact that the probability for the $k$-reaction to happen is given by $\Frac{\Sigma_k^\dagger(E')}{\Sigma_t^\dagger(E')}$. Then one can write:
\begin{eqnarray}
\label{222}
f^\dagger(E'\longrightarrow E,\hat\Omega'\cdot\hat\Omega)&=&\Frac{\Sigma_t(E)}{\Sigma_t(E')}
f(E\longrightarrow E',\hat\Omega'\cdot\hat\Omega)=\nonumber \\ && \hskip-2.5truecm
\Frac{\Sigma_t(E)}{\Sigma_t(E')}
\sum_{k=s,f,\ldots}\Frac{\Sigma_k(E)}{\Sigma_t(E)}\nu_k(E)
f_k(E\longrightarrow E',\hat\Omega'\cdot\hat\Omega)=
\\ && \hskip-2.5truecm
\sum_{k=s,f,\ldots}\Frac{\Sigma_k(E)}{\Sigma_t(E')}\nu_k(E)
f_k(E\longrightarrow E',\hat\Omega'\cdot\hat\Omega)\nonumber.
\end{eqnarray}
By equating (\ref{111}) and (\ref{222}), the following relation is established:
\begin{eqnarray}
\sum_{k=s,f,\ldots}\Frac{\Sigma_k(E)}{\Sigma_t(E')}\nu_k(E)
f_k(E\longrightarrow E',\hat\Omega'\cdot\hat\Omega)=\\\nonumber
\sum_{k=s,f,\ldots}\Frac{\Sigma_k^\dagger(E')}{\Sigma_t^\dagger(E')}\nu_k^\dagger(E')
f_k^\dagger(E'\longrightarrow E,\hat\Omega'\cdot\hat\Omega),
\end{eqnarray}
which is trivially fulfilled if for all reactions:
\begin{equation}
\Sigma_k^\dagger(E')\nu_k^\dagger(E')f_k^\dagger(E'\longrightarrow E,\hat\Omega'\cdot\hat\Omega)=
\Sigma_k(E)\nu_k(E)f_k(E\longrightarrow E',\hat\Omega'\cdot\hat\Omega).
\label{eqn:soladj}
\end{equation}
This seemingly obvious solution requires however a non trivial assumption, that pseudo-particles are subject to the same set of reactions as neutrons. This is not at all mandatory and it is simply a convenient choice for the purpose of simulation\footnote{In such a way, the data needed for the Monte Carlo simulation of pseudo-particles transport are the same as for neutrons, as they are contained, for example, in the usual nuclear data files.}. Along this line of thought we can assume that not only the total cross section for pseudo-particles is the same as for neutrons, but that the same happens for all partial reactions, that is for all $k$:
\begin{equation}
\Sigma_k^\dagger(E)=\Sigma_k(E)\,;
\end{equation}
however analogies between forward and adjoint simulation shall not go beyond this point, essentially because the true difference between the two cases is that in taking the adjoint we loose the kernel normalization (with respect to outgoing energies and directions). In fact, if we assume - as it is natural - that the partial adjoint kernels are normalized with respect to the outgoing pseudo-particle energy
\begin{equation}
\int_{4\pi}d\hat\Omega\int_0^\infty\,dE f_k^\dagger(E'\longrightarrow E,\hat\Omega'\cdot\hat\Omega)=1,
\end{equation}
by integrating (\ref{eqn:soladj}) over $E$ and $\hat\Omega$, we have:
\begin{equation}
\Sigma_k(E')\nu_k^\dagger(E')=
\int_{4\pi}d\hat\Omega\int_0^\infty\,dE\,
\Sigma_k(E)\nu_k(E)f_k(E\longrightarrow E',\hat\Omega'\cdot\hat\Omega),
\label{eqn:soladj1}
\end{equation}
which implies that, in general, the mean number of pseudo-particles outgoing from a collision is not the same as for neutrons. 

It is remarkable that with these choices neutron importance can again be interpreted as a flux (density) of pseudo-particles, so that for instance traditional collision or track-length estimators can be used throughout the simulation process: this fact is apparently in contradiction with the discussion above about the physical intepretation of the neutron importance - a dimensionless quantity - with respect to a flux - a dimensional quantity. However it should be clear that when building a transport Monte Carlo model for the solution of the importance equation we implicitly modify the meaning (not the numerical value) we attribute to the adjoint source that in this scheme really corresponds to some pseudo-particle density; in other words, we build an effective transport model for pseudo-particles whose solution - that is a flux - numerically coincides with the solution for the neutron importance, which instead is dimensionless. As a last remark, it is worth observing that the importance function for the pseudo-particles herewith introduced obeys the direct transport equation, thus establishing a full duality for the two equations, with specular physical meanings.

As an example, for the sake of simplicity, consider $s$-wave neutron scattering, for which  
$$f_k(E\longrightarrow E',\hat\Omega'\cdot\hat\Omega)=\Frac{1}{(1-\alpha)E}\eta(E,\hat\Omega,E;\hat\Omega')\theta(E'-\alpha E)\theta(E-E'),$$
where $\alpha=[(A-1)/(A+1)]^2$, $A$ being the nuclei mass number, and $\theta$ is the standard Heaviside unit step function. Here $\nu(E)=1$, as it is obvious for scattering processes. The angular function $\eta(E',\hat\Omega',E;\hat\Omega)$ is the probability density function that a particle colliding at energy $E$ and with direction $\hat\Omega$, and being emitted at energy $E'$, appears at direction $\hat\Omega'$. For the type of scattering considered, the $\eta$ function turns out to be simply related to a $\delta$- function, namely $\delta(\hat\Omega'\cdot\hat\Omega-\mu_0 (E, E'))$, where $\mu_0 (E, E')$ is the scattering angle cosine, uniquely determined by the values of $E'$ and $E$. However, for the pseudo-particles "scattering" process we have
\begin{equation}
\nu_s^\dagger(E)=\Frac{1}{(1-\alpha)\Sigma_s(E)}\int_E^{E/\alpha}\Frac{\Sigma_s(E')}{E'}dE'
\end{equation}
that is not 1 even for a constant scattering cross section $\Sigma_s(E)$ - as in such a case $\nu_s^\dagger(E)=-\ln\alpha/(1-\alpha)$:
only in the limiting case of infinite mass scatterers we recover the usual interpretation of the pseudo-scattering process. A simple calculation yields the adjoint energy-angle distribution:
\begin{equation}
 f_s^\dagger(E'\longrightarrow E,\hat\Omega'\cdot\hat\Omega)=\Frac{\Sigma_s(E)}{\int_{E'}^{E'/\alpha}\Frac{\Sigma_s(E)}{E}dE}
 \Frac{\theta(E'-\alpha E)\theta(E-E')}{E}.
\end{equation}
A similar situation occurs for pseudo-fission, where:
\begin{equation}
\nu_f^\dagger(E)=\Frac{\chi(E)}{\Sigma_f(E)}\int_0^\infty dE' \nu(E')\Sigma_f(E')
\end{equation}
and the isotropic energy-angle distribution is given by
\begin{equation}
f^\dagger_f(E'\longrightarrow E,\hat\Omega'\cdot\hat\Omega)=\Frac{\nu(E)\Sigma_f(E)}{4\pi\int_0^\infty\nu(E')\Sigma_f(E')dE'}.
\end{equation}
In both cases the energy-angle distribution are clearly normalized with respect to the outgoing pseudo-particle energy and angle.

Other definitions or choices are however possible; we mention only one interesting case: we could ask to have an adjoint energy-angle distribution of the same functional form as the original (forward) one, at the price of having a mean number of outgoing pseudo-particles which depends not only on their incoming energy, but also on the outgoing one.

In the approach presented in this work the sampling procedure for the adjoint equation proceeds as follows:
\renewcommand{\labelenumi}{(\roman{enumi})}
\begin{enumerate}
\item select a pseudo-neutron of initial weight 1 by sampling its initial energy and angle from the energy/angle distribution as it is defined by the external adjoint source distribution, or alternatively select it from a flat distribution and assign an initial weight 
$$\Frac{{\mathcal S}^\dagger(E,\hat\Omega)}{\int dE\,d\hat\Omega\; {\mathcal S}^\dagger(E,\hat\Omega)};$$
\item select which collision this pseudo-neutron undergoes on the basis of the probabilities $\Frac{\Sigma_k(E)}{\Sigma_t(E)}$;
\item select the outgoing energy and angle $(E',\hat\Omega')$ from the normalized $f_k^\dagger(E,\hat\Omega\longrightarrow E',\hat\Omega')$;
\item score appropriate estimators for the adjoint flux;
\item multiply the weight by the mean number of pseudo-particles outgoing from the collision $\nu^\dagger$;
\item if $E'$ is larger than the maximum energy of interest, terminate the history, otherwise go to (ii) with the new energy and angle from (iii);
this last point is somehow tricky if one does not select a sufficiently wide energy interval to make negligible the probability that the pseudo-neutron can in the future come back in the energy range of interest: this can happen because of pseudo-fission processes, hence the maximum energy of interest should be chosen so that $\Sigma_f(E_{max})\ll \Sigma_t(E_{max})$.
\end{enumerate}
The procedure is essentially the same as in a normal (forward) simulation, but for the last point.

\section{A neutron transport adjoint equation that allows an analytical solution}

The adjoint equation can be analytically solved in a closed form for a classical slowing down model. This solution allows obtaining valuable reference results, useful to validate the proposed adjoint Monte Carlo procedure, besides retaining a relevance per se, specially for educational purposes. 

The direct flux equation in an infinite medium whose nuclei are characterized by the mass parameter $\alpha$, for s-wave scattering and in the absence of absorption, reads:
\begin{eqnarray}
\Sigma_t(E)\phi(E)&=&S(E)+\int_0^\infty \phi(E')\Sigma_s(E'\longrightarrow E)dE',
\label{eqn:rall}
\end{eqnarray}
where 
\begin{equation}
\Sigma_s(E'\longrightarrow E)=\Frac{\Sigma_s(E')}{E'(1-\alpha)}\theta(E'- E/\alpha)\theta(E'-E).
\label{eqn:swave}
\end{equation}
Equation (\ref{eqn:rall}) is a slowing down equation because the flux at a given energy $E$ receives contributions only from fluxes at higher energies in view of Eq. (\ref{eqn:swave}), or, that is the same, since no up-scattering is present, the flux at a given energy depends on the values of the flux at higher energies.

The corresponding adjoint equation can be written in a straightforward manner as:
\begin{eqnarray}
\Sigma_s(E)\phi^\dagger(E)&=&S^\dagger(E)+\int_0^\infty \phi^\dagger(E')\Sigma_s(E\longrightarrow E')dE'.
\end{eqnarray}
Since $E'$ is the energy of a neutron {\em outgoing} from a collision event, it can contribute to the adjoint flux only at a higher energy $E$. Hence the adjoint flux at some energy depends on the adjoint flux at lower energies: consequently, while the flux equation is a {\em slowing down equation}, the adjoint equation has the opposite meaning. The equation under consideration takes then the following form:
\begin{eqnarray}
\Sigma_s(E)\phi^\dagger(E)&=&S^\dagger(E)+\Frac{\Sigma_s(E)}{E(1-\alpha)}\int_{\alpha E}^E \phi^\dagger(E')dE'.
\end{eqnarray}

Let us suppose now that $S^\dagger(E)=S_0^\dagger\delta\left(E-E_0\right)$ and that $\phi^\dagger(E)=\phi_0^\dagger\delta\left(E-E_0\right)+\phi_c^\dagger(E)$. By the equating singular parts, we immediately obtain:
$$ \Sigma_s(E)\phi_0^\dagger\delta\left(E-E_0\right)=S_0^\dagger\delta\left(E-E_0\right),$$
or
$$\phi_0^\dagger=\Frac{S_0^\dagger}{\Sigma_s(E_0)}.$$
For the adjoint collided part one gets:
\begin{eqnarray}
\phi_c^\dagger(E)&=&\Frac{1}{E(1-\alpha)}\int_{\alpha E}^E 
\left[\Frac{S_0^\dagger}{\Sigma_s(E_0)}\delta\left(E'-E_0\right)+\phi_c^\dagger(E')\right]dE'\\
&=&\Frac{\theta(E-E_0)\theta(E_0-\alpha E)}{E(1-\alpha)}\Frac{S_0^\dagger}{\Sigma_s(E_0)}
+\Frac{1}{E(1-\alpha)}\int_{\alpha E}^E \phi_c^\dagger(E') dE'.\nonumber
\end{eqnarray}
The above equation clearly admits as a solution $\phi_c^\dagger(E)=\phi_c^\dagger={\rm const}$ for $E<E_0$; the value of $\phi^\dagger_c$ can be determined by observing that no neutrons can be present for $E<E_0$, then $\phi^\dagger_c\equiv0$.
It is convenient to introduce
$f(E)=E\phi_c^\dagger(E)$ and hence:
\begin{eqnarray}
f(E)&=&\Frac{\theta(E-E_0)\theta(E_0-\alpha E)}{(1-\alpha)}\Frac{S_0^\dagger}{\Sigma_s(E_0)}
+\Frac{1}{(1-\alpha)}\int_{\alpha E}^E \Frac{f(E')}{E'} dE'\nonumber\,.
\end{eqnarray}
A source iteration process yields:
\begin{eqnarray}
f^{(0)}(E)&=&0\nonumber\\
f^{(1)}(E)&=&\Frac{\theta(E-E_0)\theta(E_0-\alpha E)}{(1-\alpha)}\Frac{S_0^\dagger}{\Sigma_s(E_0)}\nonumber\\
f^{(2)}(E)&=&\Frac{\theta(E-E_0)\theta(E_0-\alpha E)}{(1-\alpha)}\Frac{S_0^\dagger}{\Sigma_s(E_0)}+\nonumber\\ &=&
\Frac{1}{(1-\alpha)^2}\Frac{S_0^\dagger}{\Sigma_s(E_0)}\int_{\alpha E}^E \Frac{\theta(E'-E_0)\theta(E_0-\alpha E')}{E'}dE'.
\end{eqnarray}
Since necessarily $E_0<E'<E_0/\alpha$, then the last integral vanishes if $E<E_0$ or $\alpha E> E_0/\alpha$, or explicitly:
$$f^{(2)}(E)\not=0 \; \Longleftrightarrow\; E_0< E < \Frac{E_0}{\alpha^2}\,.$$
This argument can be generalized in order to write:
$$f^{(n)}(E)\not=0 \; \Longleftrightarrow\; E_0< E < \Frac{E_0}{\alpha^n}\,.$$
The presence of an overall factor $S_0^\dagger/\Sigma_s(E_0)$ clearly suggests to define $f(E)=S_0^\dagger g(E)/\Sigma_s(E_0)$, so that
\begin{eqnarray}
g(E)&=&\Frac{\theta(E-E_0)\theta(E_0-\alpha E)}{(1-\alpha)}
+\Frac{1}{(1-\alpha)}\int_{\alpha E}^E \Frac{g(E')}{E'} dE'\nonumber\,.
\end{eqnarray}
In the first interval $E_0<E<E_0/\alpha$, since $\alpha E < E_0$ and
$g(E)=0$ for $E<E_0$, the equation reads:
\begin{eqnarray}
g_1(E)&=&\Frac{1}{(1-\alpha)}
+\Frac{1}{(1-\alpha)}\int_{E_0}^E \Frac{g_1(E')}{E'} dE'\nonumber\,.
\end{eqnarray}
It is quite obvious that a solution to this equation must have the form
$$g_1(E)=\Frac{A}{E^k}$$
In fact, one can write:
\begin{eqnarray*}
\Frac{A}{E^k}&=&\Frac{1}{(1-\alpha)}+\Frac{1}{(1-\alpha)}\int_{E_0}^E\Frac{A}{y^{k+1}}dy\\ &=&
\Frac{1}{(1-\alpha)}-\Frac{1}{k(1-\alpha)}\left[\Frac{A}{E^k}-\Frac{A}{E_0^k}\right],
\end{eqnarray*}
obtaining:
\begin{eqnarray*}
\Frac{A}{E^k}=-\Frac{1}{k(1-\alpha)}\Frac{A}{E^k}\qquad &\Longrightarrow &\qquad k=-\Frac{1}{1-\alpha}\\
\Frac{1}{(1-\alpha)}+\Frac{1}{k(1-\alpha)}\Frac{A}{E_0^k}=0 &\Longrightarrow & \qquad 
A=\Frac{1}{1-\alpha}E_0^{-\Frac{1}{1-\alpha}},
\end{eqnarray*}
and, at last:
\begin{equation}
g_1(E)=\Frac{1}{1-\alpha}\left(\Frac{E}{E_0}\right)^{1/1-\alpha}.
\end{equation}

In the successive intervals $E_0/\alpha^n<E<E_0/\alpha^{n+1}$ the source term is absent because of the theta function - or because a neutron can gain a maximum fraction $1/\alpha$ of its energy for each collision it suffers, hence:
\begin{eqnarray}
g(E)&=&\Frac{1}{(1-\alpha)}\int_{\alpha E}^E \Frac{g(E')}{E'} dE'\nonumber\,.
\end{eqnarray}
The integral can be split in two terms:
\begin{eqnarray}
g_{n+1}(E)&=&\Frac{1}{(1-\alpha)}\int_{\alpha E}^{E_0/\alpha^n} \Frac{g_n(E')}{E'} dE'+\Frac{1}{(1-\alpha)}\int_{E_0/\alpha^n}^E \Frac{g_{n+1}(E')}{E'} dE',\nonumber\,.
\end{eqnarray}
so that in differential form one obtains:
\begin{equation}
\Frac{dg_{n+1}(E)}{dE}=-\Frac{1}{(1-\alpha)}\Frac{g_n(\alpha E)}{E}+\Frac{1}{(1-\alpha)}\Frac{g_{n+1}(E)}{E},
\end{equation}
with initial condition
\begin{equation}
g_{n+1}(E_0/\alpha_n)=\Frac{1}{(1-\alpha)}\int_{E_0/\alpha^{n-1}}^{E_0/\alpha^n} \Frac{g_n(E')}{E'} dE'.
\end{equation}

The above is a non homogeneous differential equation for $g_n(E)$ with a source term given by
$$S_{n+1}(E)=-\Frac{1}{(1-\alpha)}\Frac{g_n(\alpha E)}{E}\,.$$
The solution of the associated homogeneous equation is:
\begin{equation}
\Frac{dg_{n+1}^{(0)}(E)}{g_{n+1}^{(0)}(E)}=\Frac{1}{1-\alpha}\Frac{dE}{E}\qquad\Longrightarrow\qquad
g_{n+1}^{(0)}(E)=K E^\Frac{1}{1-\alpha}.
\end{equation}
By the variation of the arbitrary constant, assuming $K=K(E)$ and $g_{n+1}(E) = K(E) E^\Frac{1}{1-\alpha}$ one obtains:
\begin{equation}
\Frac{dK(E)}{dE}E^\Frac{1}{1-\alpha}=-\Frac{1}{(1-\alpha)}\Frac{g_n(\alpha E)}{E},
\end{equation}
and
\begin{eqnarray*}
K(E)&=&-\Frac{1}{(1-\alpha)}\int_{E_0/\alpha^n}^E 
\Frac{g_n(\alpha E)}{E}\Frac{1}{E^\Frac{1}{1-\alpha}} dE+Q_{n+1}
\\ &=&
-\Frac{\alpha^{\Frac{1}{1-\alpha}}}{(1-\alpha)}\int_{E_0/\alpha^{n-1}}^{\alpha E} g_n(y)\Frac{1}{y^\Frac{1}{1-\alpha}} \Frac{dy}{y}+Q_{n+1}.
\end{eqnarray*}
Recalling that $g_n$ is defined over $\left[\Frac{E_0}{\alpha^{n-1}},\Frac{E_0}{\alpha^n}\right]$ and that the maximum value allowed for $E$ in this case is $E_0/\alpha^{n+1}$, we realize that the maximum $y$ value is, correctly, $E_0/\alpha^n$. The initial condition requires
\begin{equation}
g_{n+1}(E_0/\alpha^n)=Q_{n+1}\left(\Frac{E_0}{\alpha^n}\right)^{\Frac{1}{1-\alpha}}=
\Frac{1}{(1-\alpha)}\int_{E_0/\alpha^{n-1}}^{E_0/\alpha^n} \Frac{g_n(E')}{E'} dE'
\end{equation}
and finally:
\begin{eqnarray}
g_{n+1}(E)&=&
\Frac{E^\Frac{1}{1-\alpha}}{1-\alpha}
\left[\left(\Frac{\alpha^n}{E_0}\right)^{\Frac{1}{1-\alpha}}
\int_{E_0/\alpha^{n-1}}^{E_0/\alpha^n} \Frac{g_n(E')}{E'} dE'\right.\\ && \left.\vphantom{(\Frac{a}{b})^\Frac{a}{b}}
-\alpha^{\Frac{1}{1-\alpha}}\int_{E_0/\alpha^{n-1}}^{\alpha E} g_n(y)\Frac{1}{y^\Frac{1}{1-\alpha}} \Frac{dy}{y}
\right]\,.
\nonumber
\end{eqnarray}

The solution of this equation can be given in closed form as (for a proof see Appendix \ref{sec:proof}):
\begin{eqnarray}
g_1(E)&=&\Frac{1}{1-\alpha}\left(\Frac{E}{E_0}\right)^\Frac{1}{1-\alpha}\nonumber\\
g_2(E)&=&\Frac{1}{(1-\alpha)^2}\left(\Frac{E}{E_0}\right)^\Frac{1}{1-\alpha}
\left[(1-\alpha)\left(1-\alpha^\Frac{1}{1-\alpha}\right)-\alpha^\Frac{1}{1-\alpha}\ln\Frac{\alpha E}{E_0}\right]\nonumber\\
\nonumber &\vdots& \\
g_n(E)&=&\Frac{1}{(1-\alpha)^n}\left(\Frac{E}{E_0}\right)^\Frac{1}{1-\alpha}\times\label{eqn:fullform}\\ &&
\times\left[
(1-\alpha)^{n-1}\left(1-\alpha^\Frac{1}{1-\alpha}\right)-(1-\alpha)^{n-2}\alpha^\Frac{1}{1-\alpha}\ln\Frac{\alpha E}{E_0}
+\right.\nonumber\\ && \left.\vphantom{\alpha^\Frac{1}{1-\alpha}}
\hskip-2.5truecm+\sum_{m=2}^{n-1}(-1)^{m}\alpha^\Frac{m}{1-\alpha}\left(
\Frac{(1-\alpha)^{n-m}}{(m-1)!}\ln^{m-1}\Frac{\alpha^m E }{E_0}+
\Frac{(1-\alpha)^{n-m-1}}{m!}\ln^m\Frac{\alpha^m E}{E_0}
\right)
\right],\nonumber
\end{eqnarray}
where the general expression holds for $n>2$. From this expression it is immediate to conclude that
\begin{equation}
g_n\left(\Frac{E_0}{\alpha^{n-1}}\right)=g_{n-1}\left(\Frac{E_0}{\alpha^{n-1}}\right)\qquad\qquad n>2\,.
\end{equation}
In fact the last term in the sum for $g_n$ is for $m=n-1$ and a factor $\ln\Frac{\alpha^{n-1}E}{E_0}$ is always present because the minimum power for logarithms is $m-1=n-2>0$: for 
$E=\Frac{E_0}{\alpha^{n-1}}$ this is $\ln 1=0$. The remaining terms coincide with the expression for 
$g_{n-1}\left(\Frac{E_0}{\alpha^{n-1}}\right)$.
\begin{figure}[!htb]
\centering
\includegraphics[width=0.95\linewidth]{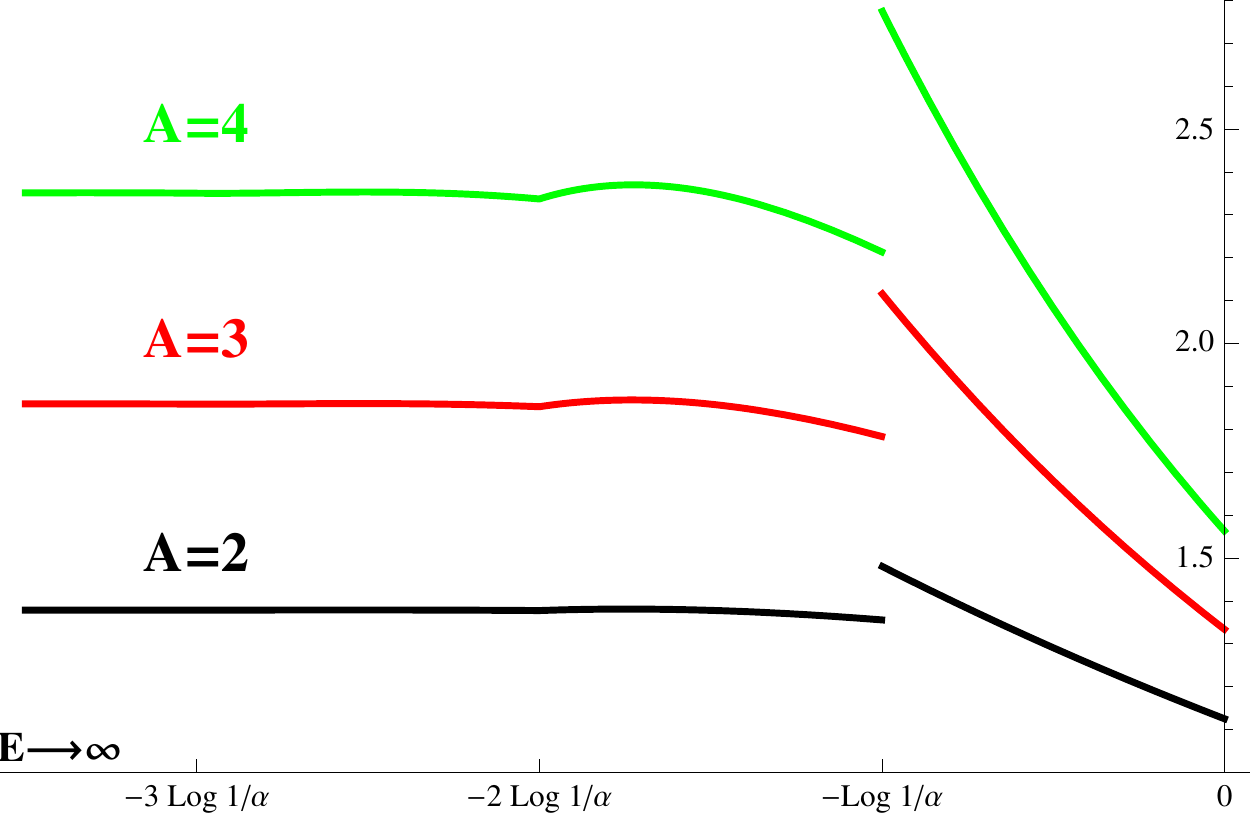}
\caption{Adjoint Placzek functions in the lethargy variable: for each plot the lethargy variable is in a different scale, in such a way discontinuities appear at the same points.}
\label{fig:AdjPlFig}
\end{figure}
The adjoint flux is given, in the corresponding energy intervals by the following formula:
\begin{equation}
\phi^\dagger_n(E)=\Frac{S^\dagger_0}{\Sigma_s(E_0) E}g_n(E)\qquad\qquad \Frac{E_0}{\alpha^{n-1}}\le E\le  \Frac{E_0}{\alpha^n},
\end{equation}
and explictly:
\begin{eqnarray}
\phi^\dagger_1(E)&=&\Frac{1}{E_0\Sigma_s(E_0)(1-\alpha)}\left(\Frac{E}{E_0}\right)^\Frac{\alpha}{1-\alpha}\nonumber\\
\phi^\dagger_2(E)&=&\Frac{1}{E_0\Sigma_s(E_0)(1-\alpha)^2}\left(\Frac{E}{E_0}\right)^\Frac{\alpha}{1-\alpha}\times\nonumber\\ &&\times
\left[(1-\alpha)\left(1-\alpha^\Frac{1}{1-\alpha}\right)-\alpha^\Frac{1}{1-\alpha}\ln\Frac{\alpha E}{E_0}\right]\nonumber\\
\nonumber &\vdots& \\
\phi^\dagger_n(E)&=&\Frac{1}{E_0\Sigma_s(E_0)(1-\alpha)^n}\left(\Frac{E}{E_0}\right)^\Frac{\alpha}{1-\alpha}\times\label{eqn:fullformadjflux}\\ &&
\times\left[
(1-\alpha)^{n-1}\left(1-\alpha^\Frac{1}{1-\alpha}\right)-(1-\alpha)^{n-2}\alpha^\Frac{1}{1-\alpha}\ln\Frac{\alpha E}{E_0}
+\right.\nonumber\\ && \left.\vphantom{\alpha^\Frac{1}{1-\alpha}}
\hskip-2.5truecm+\sum_{m=2}^{n-1}(-1)^{m-2}\alpha^\Frac{m}{1-\alpha}\left(
\Frac{(1-\alpha)^{n-m}}{(m-1)!}\ln^{m-1}\Frac{\alpha^m E }{E_0}+
\Frac{(1-\alpha)^{n-m-1}}{m!}\ln^m\Frac{\alpha^m E}{E_0}
\right)
\right].\nonumber
\end{eqnarray}
Graphs of the adjoint Placzek functions in the lethargy variable are shown in Fig.~\ref{fig:AdjPlFig}.

A comment is here useful: the reason why it is possible to obtain a compact, closed form for the adjoint flux is related to the fact that
{\em we do not make use from the beginning of the lethargy variable}
\begin{equation}
u=\ln\Frac{E_0}{E},
\end{equation}
as it is traditionally done in literature. It is quite evident from (\ref{eqn:fullformadjflux}) that it is possible to transform the above expressions in terms of lethargy variable\footnote{Alternatively the "adjoint lethargy" or "vivacity", $u^\dagger=\ln\Frac{E}{E_0}=-u$, may be used.}, but the original forms in terms of energy are much simpler. For instance:
\begin{equation}
\ln^k\Frac{\alpha^k E}{E_0}=\left(\ln\left(\alpha^k e^{-u}\right)\right)^k=\left(k\ln\alpha-u\right)^k .
\end{equation}
Anyway, in terms of the lethargy variable the discontinuities of the adjoint flux or of its derivatives occur for $u=-n\ln\Frac{1}{\alpha}$ as it is shown in the figures.
\begin{figure}[!htb]
\centering
\includegraphics[width=0.95\linewidth]{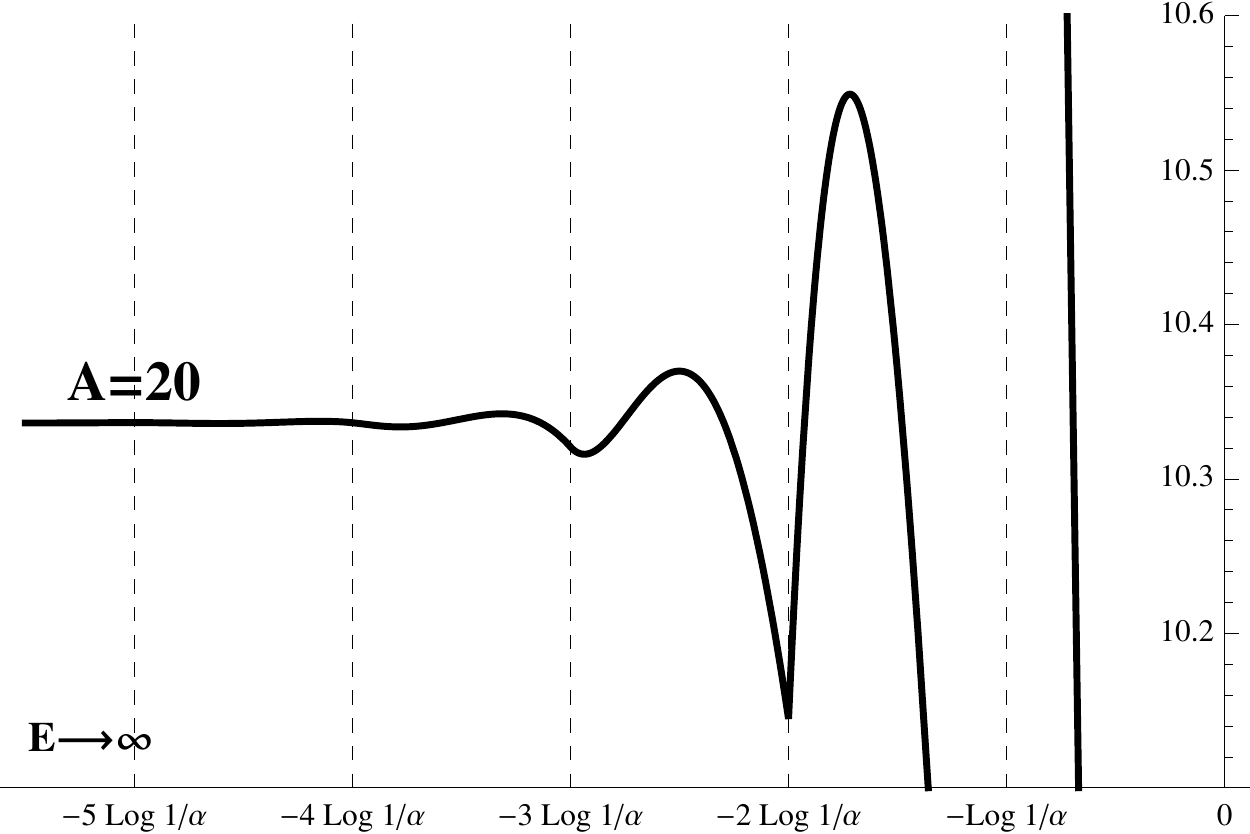}
\caption{Adjoint Placzek functions in the lethargy variable for $A=20$ to emphasize oscillations on a wider lethargy range.}
\label{fig:AdjPlFig20}
\end{figure}

In particular it can be shown that $\partial^n_u\phi^\dagger(u)$ is discontinuous at $u=-n\ln\Frac{1}{\alpha}$.
Moreover it turns out that
\begin{eqnarray*}
g_n\left(\Frac{E_0}{\alpha^n}\right)&=&
\Frac{1}{(1-\alpha)^n}\left(\Frac{1}{\alpha}\right)^\Frac{n}{1-\alpha}\times\label{eqn:finalfullform}\\ &&
\times\left[
(1-\alpha)^{n-1}\left(1-\alpha^\Frac{1}{1-\alpha}\right)-(1-\alpha)^{n-2}\alpha^\Frac{1}{1-\alpha}\ln\alpha^{1-n}
+\right.\nonumber\\ && \left.\vphantom{\alpha^\Frac{1}{1-\alpha}}
\hskip-2.5truecm+\sum_{m=2}^{n-1}(-1)^{m-2}\alpha^\Frac{m}{1-\alpha}\left(
\Frac{(1-\alpha)^{n-m}}{(m-1)!}\ln^{m-1}\alpha^{m-n}+
\Frac{(1-\alpha)^{n-m-1}}{m!}\ln^m\alpha^{m-n}\right)\right],
\end{eqnarray*}
which can be written in the more compact expression:
\begin{eqnarray}
g_n\left(\Frac{E_0}{\alpha^n}\right)&=&
\sum_{m=0}^{n-1}\Frac{(-1)^{n-m-1}}{(n-m-1)!}(1-\alpha)^{m-n}
\alpha^{-\Frac{m+1}{1-\alpha}}\times\nonumber\\ && \times
\left(\ln^{n-m-1}\Frac{1}{\alpha^{m+1}}-\alpha^\Frac{1}{1-\alpha}\ln^{n-m-1}\Frac{1}{\alpha^{m}}\right)
\label{eqn:divergent}
\end{eqnarray}
that is naturally a divergent quantity because it is the adjoint flux that has a finite limit for $E\to\infty$. This requires to divide by $E_0/\alpha^n$, so yielding the adjoint flux values at the discontinuity points as:
\begin{eqnarray}
\phi^\dagger\left(\Frac{E_0}{\alpha^n}\right)&=&
\sum_{m=0}^{n-1}\Frac{(-1)^{n-m-1}}{(n-m-1)!}(1-\alpha)^{m-n}
\alpha^{n-\Frac{m+1}{1-\alpha}}\times\nonumber\\ && \times
\left(\ln^{n-m-1}\Frac{1}{\alpha^{m+1}}-\alpha^\Frac{1}{1-\alpha}\ln^{n-m-1}\Frac{1}{\alpha^{m}}\right).
\end{eqnarray}
This is a useful expression for approximate numerical evaluations, because it is stable for moderate values of $n$. However, the finiteness of the asymptotic limit for $n\to\infty$  is guaranteed only by the factor $\alpha^n$, being the sequence defined by (\ref{eqn:divergent}) divergent.

\section{Validation of the sampling procedure}
\begin{figure}[!htb]
\centering
\includegraphics[width=0.95\linewidth]{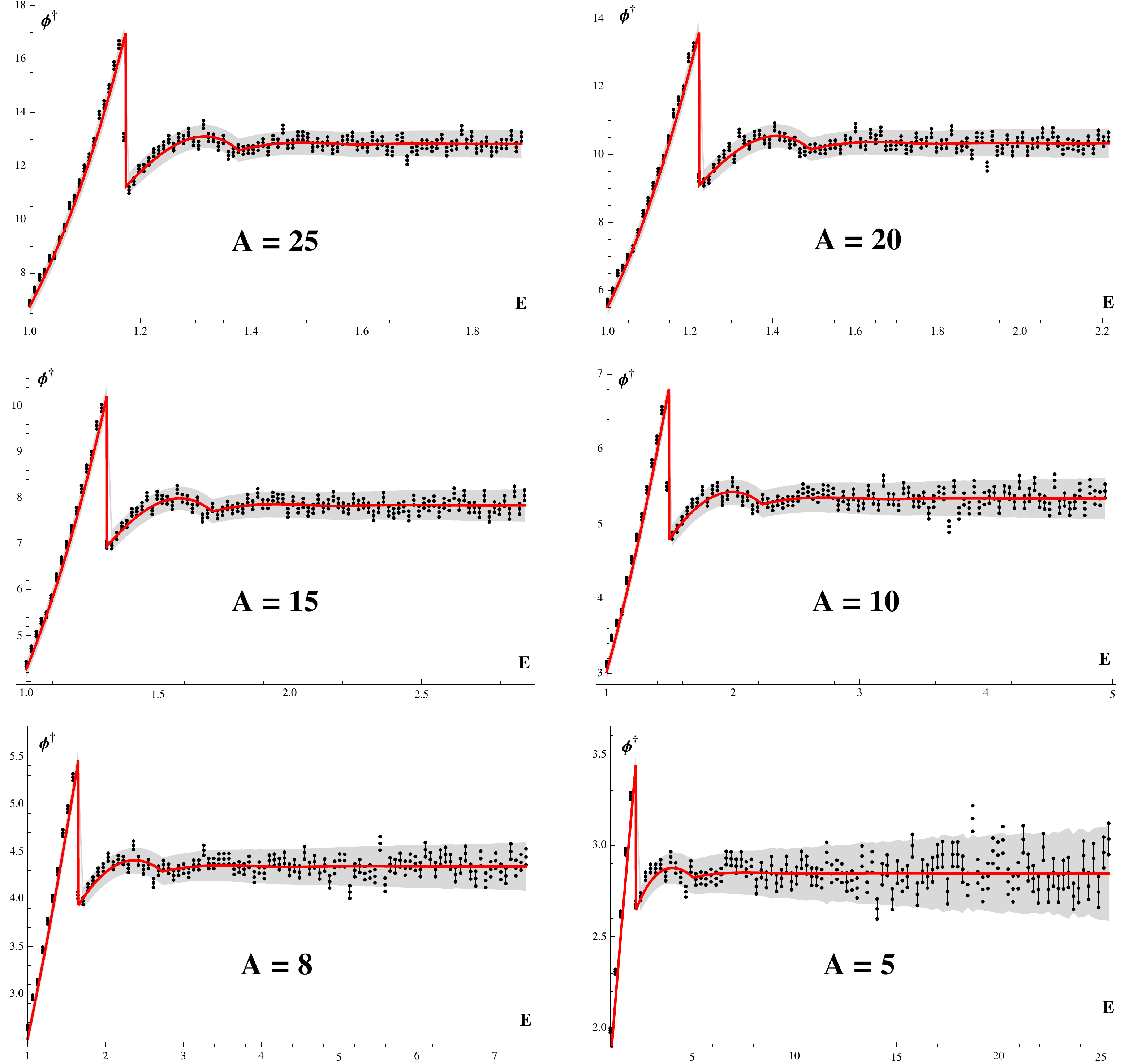}
\caption{Simulated and analytic result for the adjoint flux: for each simulated point a $1\sigma$ statistical error bar is shown; the grey area is the $3\sigma$ confidence region built around the analytical result using the estimated statistical standard deviation: as expected the mean number of points outside this confidence region does not exceed 1\%. The number of histories used in the simulation is $10^5$.}
\label{fig:AdjSimul}
\end{figure}

The implementation and test of the proposed sampling procedure for the adjoint flux appears straigthforward from the previous considerations and we show some MC simulations in Fig. \ref{fig:AdjSimul}, together with the corresponding analytic result obtained from Eq.~(\ref{eqn:fullformadjflux}).

\begin{figure}[!htb]
\centering
\includegraphics[width=0.45\linewidth]{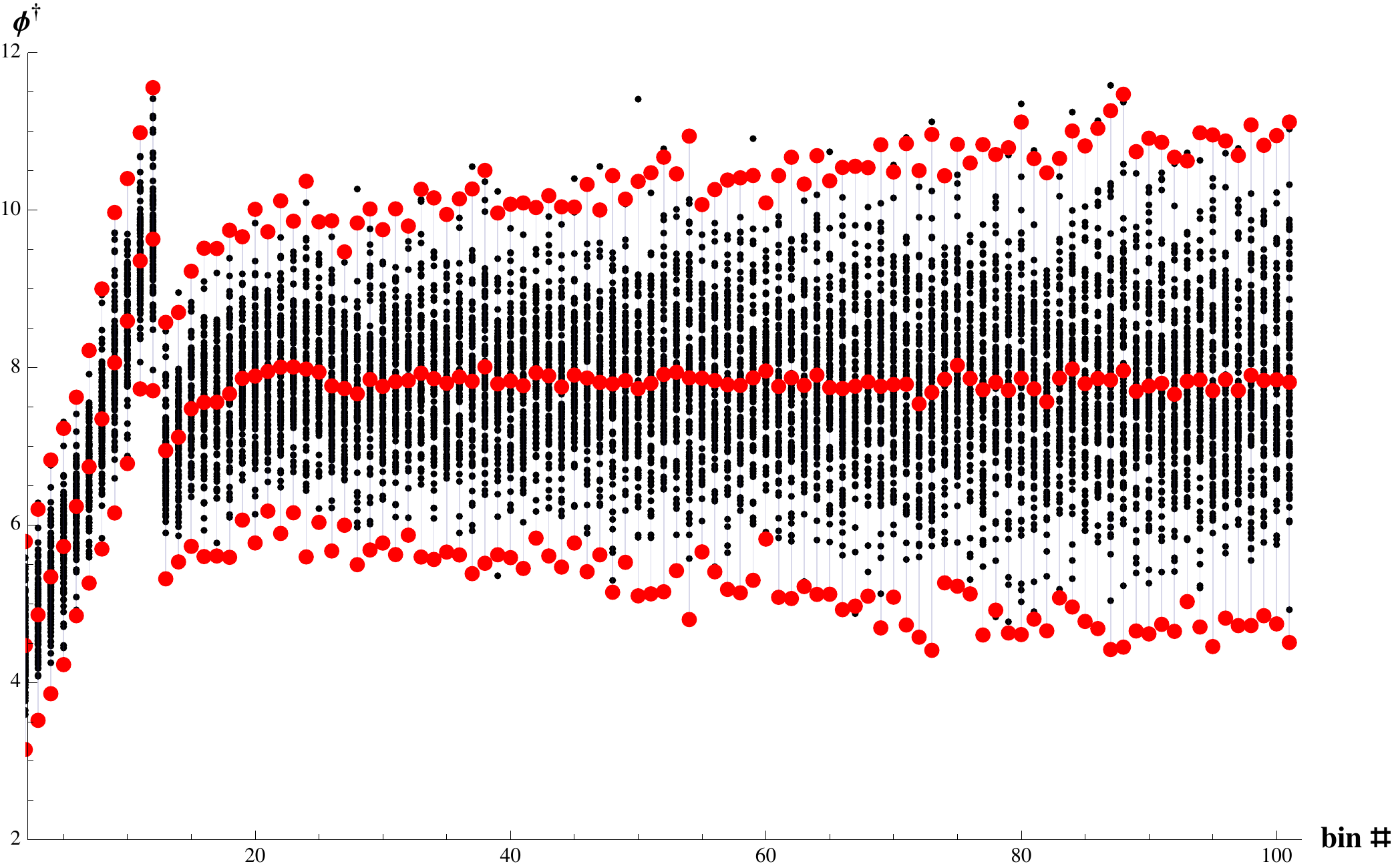}
\includegraphics[width=0.45\linewidth]{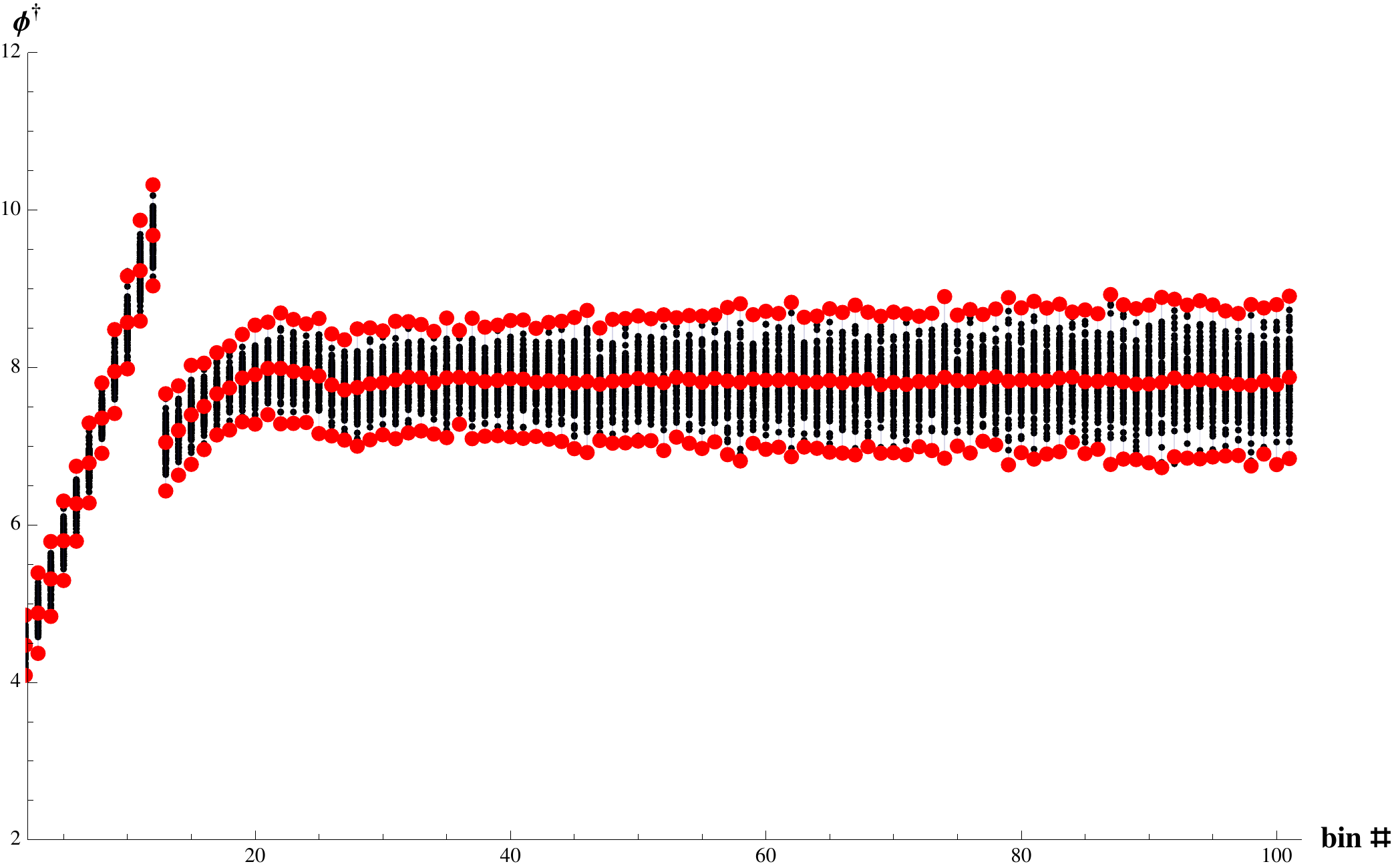}
\caption{Distribution of adjoint flux samples for $N=1000$ (on the left) and $N=10000$ (on the right): larger (red) points represent the mean of the sampled flux estimates and the lower and upper limits of respective $3\sigma$ confidence interval.  In this case $A=15$.}
\label{Fig:Sample}
\end{figure}

We have further verified that the simulation of the adjoint flux along the proposed scheme preserves all the properties that are expected by a MC sampling: in particular - as it is well known - the MC procedure provides a statistical sample of the desired quantity; this entails that the simulated adjoint flux is by itself a stochastic quantity with an associated probability distribution, with its well-defined mean and variance. For a sufficiently large number of histories $N$, the statistical error associated can be estimated as:
\begin{equation}
\sigma_{<x>}=\Frac{\sigma_x}{\sqrt N}\,.
\end{equation}
In Fig. \ref{Fig:Sample} we show that this is indeed the case: we produced a set of 100 MC estimates of the adjoint flux for $N=1000$ and $N=10000$: in each of the graphs every single adjoint flux estimate - obtained using the conventional collision estimator - is shown as a single black dot as a function of the energy bin interval.\footnote{For these simulations we used 100 uniform energy bins in the interval $(E_0,E_0/\alpha^5$) to collect results for adjoint flux estimator.}
\newline\noindent However from Figure \ref{fig:AdjSimul} we realize that the proposed simulation scheme displays an unwanted and unpleasant feature: the statistical uncertainty on the result is manifestly increasing with energy, an effect that is more pronounced for low values of the mass number $A$. This is a consequence of two concurrent effects: on one side we built simulations in the interval $E_0,E_0/\alpha^4$, with equally spaced energy bins, so that $\Delta E/E$ is not constant and this implies an increase of the statistical error. On the other side, and more and more relevant for low mass numbers, the assumed simulation scheme implies that the weight of the (adjoint) particle increases at every scattering event, again causing higher statistical errors at higher energies. While the first effect can be easily corrected using energy bins such that $\Delta E/E$ is constant, the second one requires to implement some proper variance reduction scheme, for instance splitting histories when weight becomes higher than some pre-defined cutoff.

\begin{figure}[!htb]
\centering
\includegraphics[width=0.95\linewidth]{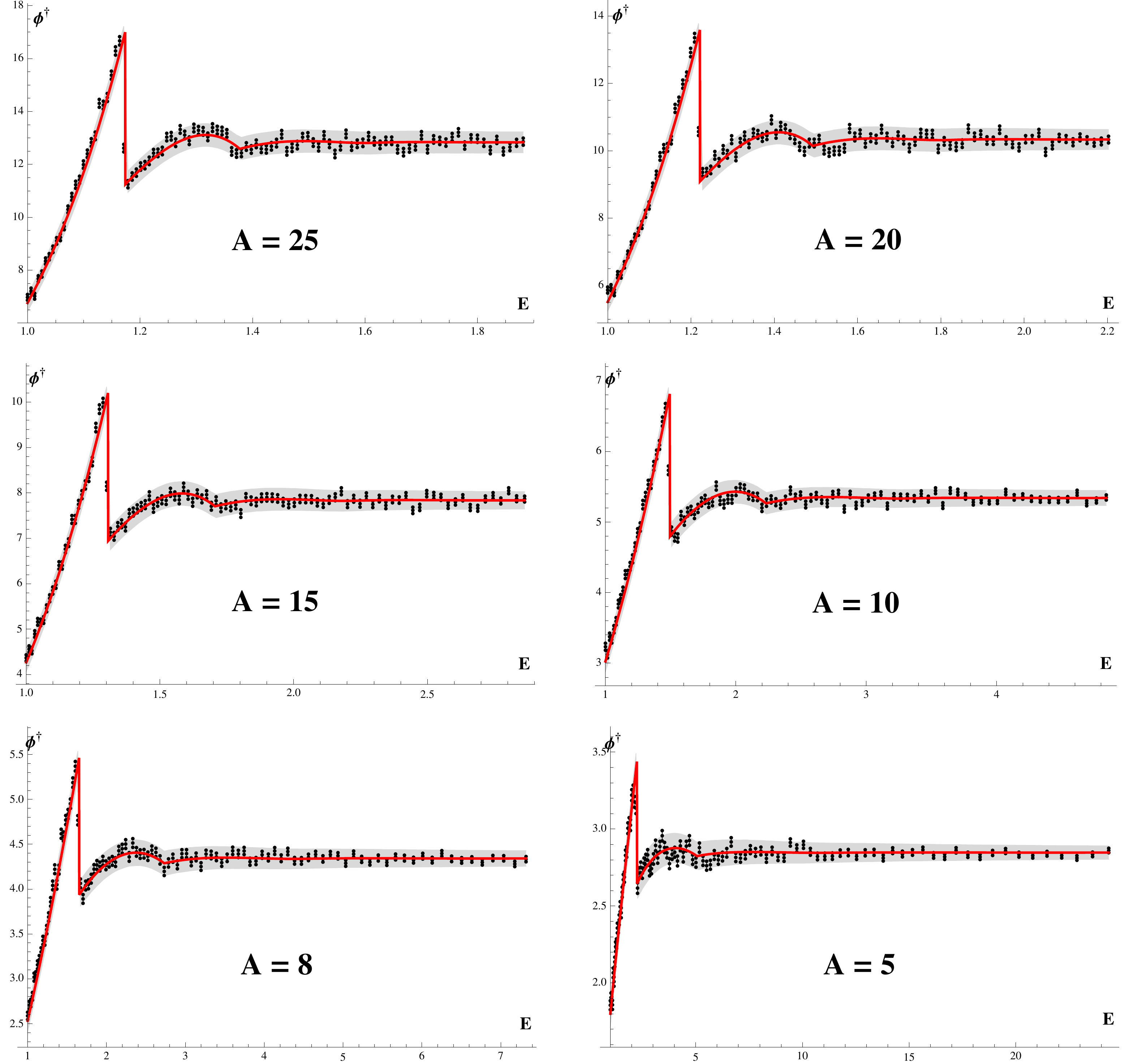}
\caption{The same as in Figure \ref{fig:AdjSimul}, but with constant $\Delta E/E$ and weight splitting enabled.}
\label{fig:AdjSimul1}
\end{figure}

We show in Figure \ref{fig:AdjSimul1} the results from simulation when both these corrections are implemented: we used weight cutoff 2, 2, 2, 2, 3 and 5 for A = 25, 20, 15, 10, 8 and 5, respectively. It is clear by comparison
of the figures \ref{fig:AdjSimul} and \ref{fig:AdjSimul1} that smoother behavior of the variance along energy variable can be obtained applying the proposed procedure.
The simulation time is enhanced by a factor of 1.07, 1.12, 1.29, 1.7, 1.65 and 2.4, respectively.

\section{Conclusions}

In this paper some basic aspects in the theory of the adjoint neutron transport equation are presented. Some important works on the physical aspects of the problem and on the relevant applications in the nuclear reactor physics field are reviewed. A sampling method enabling a Monte Carlo approach for the solution of the adjoint equation is then presented and its physical meaning is discussed in terms of the transport of virtual (adjoint) particles. The statistical procedure proposed can be applied in a straightforward manner using the same Monte Carlo tool suitable to solve direct transport problems.

In the second part of the paper a paradigmatic adjoint transport problem amenable to a fully analytical solution is considered. The solution of this problem provides a reference solution that can serve as a benchmark for the statistical procedure proposed. The problem refers to the solution of the adjoint equation for the infinite-medium slowing down process. The corresponding direct problems can be analytically solved by the use of the classic Placzek functions. In the present case, the theory of these functions is constructed and their analytical determination is carried out. This work leads to disclose novel and interesting properties of the Placzek functions, which can be established through their relationship with the adjoint ones, and to yield a full closed analytical formulation for all of them (see Appendix A).

The direct comparisons between the analytical results and those obtained by a Monte Carlo simulation allow to validate the suitability of the statistical approach proposed for the solution of the adjoint transport problem. The favorable comparison allows to conclude that the sampling procedure can be successfully applied for the determination of the adjoint flux in neutron transport for reactor physics applications.

\newpage
\appendix

\section{The Placzek functions}
It is useful and educationally worth-while to see how the previous technique applies also to the calculation of the "original" Placzek functions. It is known that they obey the following equation \cite{duder}:
\begin{equation}
\Frac{dF(E)}{dE}=\Frac{F\left(\Frac{E}{\alpha}\right)}
{(1-\alpha)\Frac{E}{\alpha}}-\Frac{F(E)}{(1-\alpha)E}\,.
\end{equation}
They are discontinuous or they have discontinuous derivatives at $E=\alpha^n E_0$, so they are appropriately defined as $F_n$ over intervals $\alpha^n E_0<E\le \alpha^{n-1} E_0$. The first two of these functions, in terms of the lethargy variable are given by (cfr. {\em ibidem} eqn. (8-50) and (8-55)):
\begin{eqnarray}
F_1(u) &=& S_0 \Frac{\exp\left[\Frac{\alpha}{1-\alpha}u\right]}{1-\alpha}\\
F_2(u) &=& S_0\left(\Frac{1-\alpha^\Frac{1}{1-\alpha}}{1-\alpha}\right)\exp\left[\Frac{\alpha}{1-\alpha}u\right]
-\\ && -S_0 \Frac{\alpha^\Frac{\alpha}{1-\alpha}}{(1-\alpha)^2}\left(u-\ln\Frac{1}{\alpha}\right)
\exp\left[\Frac{\alpha}{1-\alpha}u\right],
\end{eqnarray}
which can be translated into the energy variable as \footnote{It must be recalled that the definition given for the lethargy dependent collision density is such that $$F(u)du=-F(E)dE,$$ which entails $$F(u)=E F(E).$$}:
\begin{eqnarray}
F_1(E) &=& \Frac{S_0\,\left(\Frac{E_0}{E}\right)^\Frac{\alpha}{1-\alpha}}{E(1-\alpha)}\nonumber\\
F_2(E) &=& \Frac{S_0\,\left(\Frac{E_0}{E}\right)^\Frac{\alpha}{1-\alpha}}{E(1-\alpha)^2}
\left[(1-\alpha)\left(1-\alpha^\Frac{1}{1-\alpha}\right)-
\right.\\ && \nonumber\left.\vphantom{\left(\alpha^\Frac{1}{1-\alpha}\right)}
-\alpha^\Frac{\alpha}{1-\alpha}\ln\Frac{\alpha E_0}{E}\right].
\end{eqnarray}
It is remarkable that in these two intervals ($\alpha^2 E_0 < E < \alpha E_0$ and $\alpha E_0 <  E < E_0$) the following property holds:
\begin{eqnarray}
\Frac{E_0\Sigma_s(E_0)}{S_0^\dagger}\phi^\dagger_{1/2}(E)&=&h_{1/2}\left(\Frac{E}{E_0}\right),\nonumber\\
\Frac{E\Sigma_s(E)}{S_0}\phi_{1/2}(E)&=&h_{1/2}\left(\Frac{E_0}{E}\right).
\label{eqn:firsttwo}
\end{eqnarray}

Let us suppose now that a function exists such that the previous relation hold over all the allowed energy range, that is:
\begin{eqnarray*}
\Frac{E_0\Sigma_s(E_0)}{S_0^\dagger}\phi^\dagger(E)&=&h\left(\Frac{E}{E_0}\right),\\
\Frac{E\Sigma_s(E)}{S_0}\phi(E)&=&h\left(\Frac{E_0}{E}\right).
\end{eqnarray*}
To prove the consistency of this hypothesis we can start from the equations for $\phi^\dagger$ and for $\Sigma_S(E)\phi(E)$
and show that they imply the same equation for $h(y)$\footnote{In the case of ordinary first order differential equations to prove that two functions are identical it is sufficient to prove that they obey to the same equation and that they coincide in one point;
this case is a little bit more complicated because differential equations involved in the game are not ordinary ones, but first order differential-difference equations in lethargy variable. In this case to fully specify a solution we must specify its values on the interval
$1<x<1/\alpha$ for the argument of $g(x)$.}: this is a necessary and sufficient condition because we have shown
by inspection that (\ref{eqn:firsttwo}) holds in the first two respective intervals of energy\footnote{Effectively, the transformation
$\Frac{E}{E_0}\longleftrightarrow\Frac{E_0}{E}$ maps different energy intervals on a single one for the function $h(x)$.}.

\noindent
The equation to be satisfied by $\phi^\dagger$ for $E>E_0/\alpha$ is 
\begin{eqnarray*}
\Frac{d\phi^\dagger(E)}{dE}&=&\Frac{\alpha}{E(1-\alpha)}\phi^\dagger(E)-\Frac{\alpha }{E(1-\alpha)}\phi^\dagger(\alpha E)\\ 
\end{eqnarray*}
and, multiplying by $\Frac{E_0\Sigma_s(E_0)}{S^\dagger_0}$,
\begin{eqnarray}
\Frac{dh\left(\Frac{E}{E_0}\right)}{dE}&=&\Frac{\alpha}{E(1-\alpha)}h\left(\Frac{E}{E_0}\right)
-\Frac{\alpha }{E(1-\alpha)}h\left(\Frac{\alpha E}{E_0}\right)\;. 
\label{eqn:A6}
\end{eqnarray}
On the other hand the equation satisfied by $F_c(E)=\Sigma_s(E)\phi(E)$ is \cite{duder} for $E<\alpha E_0$:
\begin{equation*}
\Frac{dF_c(E)}{dE}=\Frac{1}{E(1-\alpha)}\left[F_c\left(\Frac{E}{\alpha}\right)-F_c(E)\right]
\end{equation*}
and then 
\begin{eqnarray*}
\Frac{d}{dE}\Frac{E}{S_0}F_c(E)&=&\Frac{1}{S_0}F_c(E)+\Frac{E}{S_0}\Frac{dF_c}{dE}=
\Frac{1}{S_0}F_c(E)+\Frac{E}{S_0}\Frac{1}{E(1-\alpha)}\left[F_c\left(\Frac{E}{\alpha}\right)-F_c(E)\right]\\
 &=&\Frac{1}{S_0}F_c(E)\left[1-\Frac{1}{1-\alpha}\right]+\Frac{1}{S_0}\Frac{1}{(1-\alpha)}F_c\left(\Frac{E}{\alpha}\right)\\
&=&\Frac{1}{S_0}\Frac{1}{(1-\alpha)}F_c\left(\Frac{E}{\alpha}\right)-\Frac{1}{S_0}\Frac{\alpha}{1-\alpha}F_c\left(E\right),
\end{eqnarray*}
or
\begin{eqnarray*}
\Frac{d}{dE}h\left(\Frac{E_0}{E}\right)&=&\Frac{\alpha}{E(1-\alpha)}h\left(\Frac{\alpha E_0}{E}\right)-
\Frac{\alpha}{E(1-\alpha)}h\left(\Frac{E_0}{E}\right).
\end{eqnarray*}
Now if we let $y=\Frac{E_0}{E}$, we have: 
\begin{eqnarray}
-\Frac{E_0}{E^2}\Frac{d}{dy}h\left(y\right)&=&\Frac{\alpha}{E(1-\alpha)}h\left(\alpha y\right)-
\Frac{\alpha}{E(1-\alpha)}h\left(y\right),
\end{eqnarray}
or
\begin{equation}
\Frac{d}{dy}h\left(y\right)=\Frac{\alpha}{y(1-\alpha)}h\left(y\right)-\Frac{\alpha}{y(1-\alpha)}h\left(\alpha y\right)\qquad y>1/\alpha\,.
\label{eqn:A7}
\end{equation}
On the other hand in (\ref{eqn:A6}) we can substitute $z=E/E_0$ (and again we are constrained to $z>1/\alpha$),
obtaining
\begin{eqnarray}
\Frac{dh\left(z\right)}{dz}&=&\Frac{\alpha}{z(1-\alpha)}h\left(z\right)
-\Frac{\alpha }{z(1-\alpha)}h\left(\alpha z\right)\;. 
\end{eqnarray}
which is manifestly the same equation as (\ref{eqn:A7}):
\section{\label{sec:proof} Proof of equation (\ref{eqn:fullform})}
Here we give the proof of eqn. (\ref{eqn:fullform}):
\begin{eqnarray}
g_1(E)&=&\Frac{1}{1-\alpha}\left(\Frac{E}{E_0}\right)^\Frac{1}{1-\alpha}\nonumber\\
g_2(E)&=&\Frac{1}{(1-\alpha)^2}\left(\Frac{E}{E_0}\right)^\Frac{1}{1-\alpha}
\left[(1-\alpha)\left(1-\alpha^\Frac{1}{1-\alpha}\right)-\alpha^\Frac{1}{1-\alpha}\ln\Frac{\alpha E}{E_0}\right]\nonumber\\
g_n(E)&=&\Frac{1}{(1-\alpha)^n}\left(\Frac{E}{E_0}\right)^\Frac{1}{1-\alpha}\times\label{eqn:fullformA}\\ &&
\times\left[
(1-\alpha)^{n-1}\left(1-\alpha^\Frac{1}{1-\alpha}\right)-(1-\alpha)^{n-2}\alpha^\Frac{1}{1-\alpha}\ln\Frac{\alpha E}{E_0}
+\right.\nonumber\\ && \left.\vphantom{\alpha^\Frac{1}{1-\alpha}}
\hskip-2.5truecm+\sum_{m=2}^{n-1}(-1)^{m}\alpha^\Frac{m}{1-\alpha}\left(
\Frac{(1-\alpha)^{n-m}}{(m-1)!}\ln^{m-1}\Frac{\alpha^m E }{E_0}+
\Frac{(1-\alpha)^{n-m-1}}{m!}\ln^m\Frac{\alpha^m E}{E_0}
\right)
\right].\nonumber
\end{eqnarray}
The term $g_2(E)$ can be found carrying out the following steps:
\begin{eqnarray*}
g_2(E)&=&\Frac{E^\Frac{1}{1-\alpha}}{1-\alpha}\left[\left(\Frac{\alpha}{E_0}\right)^\Frac{1}{1-\alpha}
\int_{E_0}^{E_0/\alpha}\Frac{g_1(y)}{y}dy-\alpha^\Frac{1}{1-\alpha}\int_{E_0}^{\alpha E}\Frac{g_1(y)}{y^\Frac{1}{1-\alpha}}\Frac{dy}{y}
\right]\\
&=&\Frac{\left(\Frac{E}{E_0}\right)^\Frac{1}{1-\alpha}}{(1-\alpha)^2}\left[\left(\Frac{\alpha}{E_0}\right)^\Frac{1}{1-\alpha}
\int_{E_0}^{E_0/\alpha}y^{\Frac{1}{1-\alpha}-1}dy-\alpha^\Frac{1}{1-\alpha}\int_{E_0}^{\alpha E}\Frac{dy}{y}
\right]\\
&=&\Frac{\left(\Frac{E}{E_0}\right)^\Frac{1}{1-\alpha}}{(1-\alpha)^2}\left[\left(\Frac{\alpha}{E_0}\right)^\Frac{1}{1-\alpha}
(1-\alpha)\left.y^\Frac{1}{1-\alpha}\right\vert_{E_0}^{E_0/\alpha}
-\alpha^\Frac{1}{1-\alpha}\ln\Frac{\alpha E}{E_0}
\right]\\
&=&\Frac{\left(\Frac{E}{E_0}\right)^\Frac{1}{1-\alpha}}{(1-\alpha)^2}\left[
(1-\alpha)\left(1-\alpha^\Frac{1}{1-\alpha}\right)
-\alpha^\Frac{1}{1-\alpha}\ln\Frac{\alpha E}{E_0}
\right].
\end{eqnarray*}
Then we recognize that (\ref{eqn:fullformA}) implies for $n>2$ an alternative recurrence relation for the functions $g_n(E)$, namely:
\begin{eqnarray}
g_n(E)&=&g_{n-1}(E)+\Frac{(-1)^{n-1}}{(1-\alpha)^n}\left(\Frac{E}{E_0}\right)^\Frac{1}{1-\alpha}
\alpha^\Frac{n-1}{1-\alpha}\times\label{eqn:altformA}\\ \nonumber &&\times\left(
\Frac{(1-\alpha)}{(n-2)!}\ln^{n-2}\Frac{\alpha^{n-1} E }{E_0}+
\Frac{1}{(n-1)!}\ln^{n-1}\Frac{\alpha^{n-1} E}{E_0}
\right)\\ \nonumber &\equiv& g_{n-1}(E)+\Delta_n(E),
\end{eqnarray}
from which it is also immediate to conclude that
\begin{equation}
g_n\left(\Frac{E_0}{\alpha^{n-1}}\right)=g_{n-1}\left(\Frac{E_0}{\alpha^{n-1}}\right).
\end{equation}
Form (\ref{eqn:altformA}), which is easily seen as perfectly equivalent to (\ref{eqn:fullformA})\footnote{Because one simply makes explicit the last term in the sum and simplifies an overall factor $1-\alpha$ in the remaining terms, so reproducing the same form (\ref{eqn:fullformA}), with $n$ replaced by $n-1$.}, is simpler to use to obtain a proof by mathematical induction. Suppose in fact that the equation defining $g_n$:
\begin{equation}
\Frac{dg_n(E)}{dE}=-\Frac{1}{1-\alpha}\Frac{g_{n-1}(\alpha E)}{E}+\Frac{1}{1-\alpha}\Frac{g_{n}(E)}{E}
\end{equation}
holds for some value of $n>2$. Next insert (\ref{eqn:altformA}) for $g_{n+1}(E)$; we have
\begin{eqnarray*}
\Frac{dg_{n+1}(E)}{dE}&=&\Frac{dg_n(E)}{dE}+\Frac{d\Delta_{n+1}(E)}{dE}.
\end{eqnarray*}
We must verify that this expression is equal to the following one:
\begin{eqnarray*}
&&-\Frac{1}{1-\alpha}\Frac{g_{n}(\alpha E)}{E}+\Frac{1}{1-\alpha}\Frac{g_{n+1}(E)}{E}\\
&&-\Frac{1}{1-\alpha}\Frac{g_{n-1}(\alpha E)+\Delta_n(\alpha E)}{E}
+\Frac{1}{1-\alpha}\Frac{g_{n}(E)+\Delta_{n+1}(E)}{E},
\end{eqnarray*}
that is to say that $\Delta_n(E)$ itself satisfies the same equation as the $g_n$'s; however, by definition the following equality holds:
\begin{eqnarray*}
\Frac{d\Delta_n(E)}{dE}=\Frac{\Delta_n(E)}{E(1-\alpha)}+\Frac{(-1)^{n-1}}{(1-\alpha)^n}\left(\Frac{E}{E_0}\right)^\Frac{1}{1-\alpha}
\alpha^\Frac{n-1}{1-\alpha}\Frac{d}{dE}\left(\cdots\vphantom{\alpha^\Frac{a}{b}}\right).
\end{eqnarray*}
Therefore, to prove our thesis we must simply show that
\begin{eqnarray*}
-\Frac{\Delta_{n-1}(\alpha E)}{(1-\alpha)E}&=&
\Frac{(-1)^{n-1}}{(1-\alpha)^n}\left(\Frac{E}{E_0}\right)^\Frac{1}{1-\alpha}
\alpha^\Frac{n-1}{1-\alpha}\times\\ &&\times\Frac{d}{dE}\left(
\Frac{(1-\alpha)}{(n-2)!}\ln^{n-2}\Frac{\alpha^{n-1} E }{E_0}+
\Frac{1}{(n-1)!}\ln^{n-1}\Frac{\alpha^{n-1} E}{E_0}
\right) \\ &=&\Frac{(-1)^{n-1}}{(1-\alpha)^n}\left(\Frac{\alpha E}{E_0}\right)^\Frac{1}{1-\alpha}
\alpha^\Frac{n-2}{1-\alpha}\Frac{1}{E}\times\\ &&
\left(
\Frac{(1-\alpha)}{(n-3)!}\ln^{n-3}\Frac{\alpha^{n-1} E }{E_0}+
\Frac{1}{(n-2)!}\ln^{n-2}\Frac{\alpha^{n-1} E}{E_0}
\right),
\end{eqnarray*}
or
\begin{eqnarray*}
\Delta_{n-1}(\alpha E)&=&
\Frac{(-1)^{n-2}}{(1-\alpha)^{n-1}}\left(\Frac{\alpha E}{E_0}\right)^\Frac{1}{1-\alpha}
\alpha^\Frac{n-2}{1-\alpha}\times\\ &&
\times\left(
\Frac{(1-\alpha)}{(n-3)!}\ln^{n-3}\Frac{\alpha^{n-2} (\alpha E) }{E_0}+
\Frac{1}{(n-2)!}\ln^{n-2}\Frac{\alpha^{n-2} (\alpha E)}{E_0}
\right),
\end{eqnarray*}
which, being
\begin{eqnarray*}
\Delta_n(E) &=&\Frac{(-1)^{n-1}}{(1-\alpha)^n}\left(\Frac{E}{E_0}\right)^\Frac{1}{1-\alpha}
\alpha^\Frac{n-1}{1-\alpha}\times\\ \nonumber &&\times\left(
\Frac{(1-\alpha)}{(n-2)!}\ln^{n-2}\Frac{\alpha^{n-1} E }{E_0}+
\Frac{1}{(n-1)!}\ln^{n-1}\Frac{\alpha^{n-1} E}{E_0}\right)\,,
\end{eqnarray*}
is trivially verified.\hfill QED

\newpage

\end{document}